\documentclass[epj,final]{svjour}

\usepackage{amsmath}
\usepackage{amsfonts} 
\usepackage{graphicx}
\usepackage{rotating}
\usepackage{amssymb}
\usepackage{mathptmx}
\usepackage[numbers]{natbib}

\usepackage[normalem]{ulem}
\usepackage{color}

\begin{document}

\title{Ground state properties and excitation spectrum of a two dimensional gas of 
bosonic dipoles}

\author{A. Macia, F. Mazzanti, J. Boronat}

\institute{Departament de F\'{\i}sica i Enginyeria Nuclear, Campus Nord
  B4-B5, Universitat Polit\`ecnica de Catalunya, E-08034 Barcelona, Spain
  \\
  \email{adrian.macia@upc.edu}
}

\date{\today}


\abstract{
  We present a quantum Monte Carlo study 
  of two-dimensional dipolar Bose gases in the limit of zero temperature.
  The analysis is mainly focused on the anisotropy effects induced in the 
  homogeneous gas when the polarization angle with respect to the plane is
  changed. We restrict our study to the regime where the dipolar interaction is 
  strictly repulsive, although the strength of the pair repulsion depends on the
  vector interparticle distance.  
  Our results show that the effect of the anisotropy in the energy per
  particle scales with the gas parameter at low densities as expected, and that
  this scaling is preserved for all polarization angles even at the largest
  densities considered here.
  We also evaluate the excitation spectrum of the dipolar Bose gas in the context of 
  the Feynman
  approximation and compare the results obtained with the Bogoliubov ones. 
  As expected, we find that these two approximations agree at very 
  low densities, while they start to deviate from each other as the density increases. 
  For the largest densities studied, we observe a significant influence of
  the anisotropy of the dipole-dipole interaction in the excitation
  spectrum. 
 }
 


\maketitle

\section{Introduction}

In 2005 Griesmaier \textit{et al} \cite{Griesmaier_2005} and Stuhler \textit{et al} 
\cite{Stuhler_2005} obtained the first Bose-Einstein condensation of $^{52}$Cr atoms,
an achievement that 
triggered the advent of a lot of new theoretical and experimental work on 
dipolar Bose-Einstein
condensates (DBEC). The main difference between $^{52}$Cr condensates and previous 
ones with other alkalis
is the moderately 
large magnetic dipolar momentum of Chromium ($6\mu_B$)
which makes dipolar interaction comparable in strength to the usual Van der Waals 
forces, thus producing significant new
effects that can be measured in experiments. 

Dipolar quantum gases are interesting from the experimental and theoretical points 
of view due to the novel 
features  that the 
dipolar interaction introduces: anisotropy 
and long range. Dipolar interaction can be either attractive or repulsive depending 
on the relative orientation of
both the position and the dipolar moments 
of the particles, a fact that can crucially affect the stability of the 
many-body system and introduces a new degree of freedom which enriches the 
phase diagram. On the other hand, the long range character of the 
dipolar interaction leads to scattering properties 
that are radically different from
those found on the usual short-ranged potentials of quantum gases.

On the experimental side there are new exciting results coming from the work 
with polar molecules on one side\cite{KKNi_2008, Ospelkaus_2009, KKNi_2010} and 
on Dysprosium~\cite{Lu_2011, Lu_2012} and Erbium~\cite{Aikawa_2012} 
condensates on the other. Polar molecules have a large and tunable electric dipole 
moment, but 
are difficult to cool down to quantum degeneracy while keeping the system stable.
On the other hand, Dysprosium and Erbium have magnetic moments comparable to 
$^{52}$Cr but a significantly larger mass, leading to a 
much larger dipolar coupling constant with the added benefits of producing a neat
and stable magnetic dipolar condensate where dipolar effects 
are more important than the Van der Waals forces.

Dipolar quantum gases are also challenging from the theoretical point of 
view. Many works studying several characteristics of these systems have
been published in the last years. 
It is known that an homogeneous dipolar quantum gas is dynamically unstable 
against collapse in three dimensions, 
while the trapped case is conditionally stable depending 
on the geometry of the trapping potential \cite{Santos_2000,Santos_2003}. 
This implies that the DBEC stability is 
enhanced in pancake-like traps, 
where the
dipolar interaction is globally repulsive.
Low-dimensional dipolar gases have also gathered 
major theoretical interest
because the typical experimental setup involves strongly anisotropic traps 
to stabilize the system. In particular, these trapping potentials 
can be tight enough to make the system effectively
two- or one-dimensional. Many interesting studies concerning 
two- or quasi-two dimensional dipolar quantum gases have been performed 
in recent years, including the
analysis of 
two-body scattering properties 
\cite{Ticknor_2009,Ticknor_2010,Ticknor_2011}, 
static properties of the many body trapped system~\cite{Santos_2000,Cai_2010} and 
homogeneous  gas
~\cite{Buchler_2007,Astra_2007,Astra_2009,Filinov_2010,Macia_2011}, 
and some works about the dynamic response
~\cite{Mazzanti_2009,Ticknor_2010b,Hufnagl_2011}.

\begin{figure}[b]
  \begin{center}
    \includegraphics[width=0.8\linewidth]{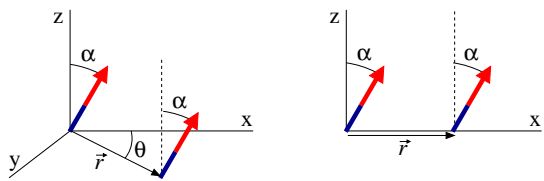}
  \end{center}
  \caption{(Color online) \textit{Left:} Two dipoles confined to move on the X-Y
    plane. The polarization field lays on the XZ plane and 
    fixes a direction in space forming an angle $\alpha$ with the $z$ axis.
    $r$ and $\theta$ are  polar coordinates in the X-Y plane.
    \textit{Right:} Particular case of two dipoles on the X-Z plane
    ($y=0$).}
  \label{petardosdeferia}
\end{figure}

In this 
work, we analyze the two-dimensional quantum gas of 
fully polarized
bosonic dipoles.
The dipolar interaction, $V_{dd}(\textbf{r})$, between two dipoles is given by
\begin{equation}
  V_{\text{dd}}(\textbf{r}) = \frac{C_{\text{dd}}}{4\pi}\left[
    \frac{\hat{\textbf{p}}_1\cdot\hat{\textbf{p}}_2-3(\hat{\textbf{p}}_1\cdot
    \hat{\textbf{r}})
      (\hat{\textbf{p}}_2\cdot\hat{\textbf{r}})}{r^3} \right]   \ ,
  \label{dipdipV}
\end{equation} 
with $\textbf{r}$ the relative position vector between them and $C_{dd}$ the 
constant defining the strength of the dipolar
interaction. For permanent magnetic dipoles, $C_{dd} = \mu_0 \mu^2$ where
$\mu_0$ is the permeability of vacuum and $\mu$ is the dipole moment of the atoms. 
Alternatively,
the electric dipole moment can be induced by an electric field $E$, and in
this case the coupling constant is $C_{dd} = d^2/\epsilon_0$, where $d = \tilde{\alpha} E$ 
with $\tilde{\alpha}$ the static polarizability and $\epsilon_0$ the permittivity of vacuum.
For a system of fully polarized dipoles in 2D as the ones considered in this work, 
$\textbf{p}_1$ and $\textbf{p}_2$ are parallel
to the polarization field (which lays on the xz-plane), and form 
an angle $\alpha$ with the normal direction to the plane, defining
a fixed direction in 
space, see Fig. \ref{petardosdeferia}. In this case $V_{dd} (\textbf{r})$ simplifies to
\begin{equation}
  V_{\text{dd}}(r,\theta) =
  \frac{C_{\text{dd}}}{4\pi}\frac{1-3\lambda^2\cos^2\theta}{r^3} \ ,
  \label{DD_interaction}
\end{equation}
where $\lambda = \sin \alpha$ and $(r,\theta)$ the
in-plane distance and polar angle, respectively. Notice 
that, since $\alpha$ is fixed, 
$\lambda \le 1$ is a constant of the problem. As usual in the study of dipolar gases, we express all quantities in 
dipolar units, dipolar length and energy, which are given by 
$r_d = mC_{dd} /4\pi \hbar^2$ and $\epsilon_d = \hbar^2 /mr_d^2$, respectively.
Another important feature of the dipolar interaction in two dimensions is that, 
contrarily to the 3D case, it is  short ranged. 
We restrict our study to polarization angles where the interaction is fully
repulsive, i. e. $\alpha \le \alpha_c \approx 0.615$, where the stability of 
the system against collapse is ensured but the interaction may still 
show a high 
degree of anisotropy. 

By means of diffusion Monte Carlo (DMC) simulations we evaluate the equation of 
state of the system, extending previous results \cite{Macia_2011} up to values 
of the density well above the 
mean-field regime, and show the effects of the 
anisotropy in the energy per particle 
and in some of the most relevant ground-state
structural quantities of the system. 
We also 
present the excitation spectrum of the anisotropic gas in both the 
Bogoliubov and Feynman
\cite{Feynman_1954} 
approximations. 
We compare these two approximations at low density, showing 
that they both agree well with each other. At higher densities, the
Bogoliubov approximation breaks down, so we use the Feynman approximation
to show how a roton minimum emerges and develops differently as 
a function of the direction in momentum space.

The rest of the 
paper is organized as follows. In 
Section II, we present DMC results 
for the ground state of the 
2D gas of tilted dipoles. The effects of the anisotropic
interaction 
in both the energy and structure properties when the 
density increases are discussed. In Section III, we calculate the 
excitation spectrum using the Feynman 
approximation, relying on the DMC results for the static structure functions,
and compare it with the Bogoliubov spectrum. Also in this case, relevant 
signatures
of the anisotropy are observed, mainly around the roton momenta. Finally,
in Section IV we present the summary and conclusions 
of our work.

\section{Ground state: energy and structure}

We have studied the many-body properties of a two-dimensional gas of
bosonic dipoles using the DMC method. DMC is a zero-temperature 
first-principles stochastic method which leads to exact properties of the 
ground state of bosonic systems. It is a form of Green's Function Monte Carlo
which samples the projection of the ground state from the initial configuration
with the operator $\exp{\left[-(\mathcal{H}-E_0)\tau\right]}$. Here, $\mathcal{H}$
is the Hamiltonian of the system, $E_0$ is a norm-preserving adjustable
constant, 
and $\tau$
is the variable which corresponds to imaginary time. The simulation is 
performed by evolving in  time $\tau$ by means of a combination of diffusion, drift and 
branching
steps acting on walkers (sets of $2N$ coordinates) representing the wavefunction of 
the system.

The Hamiltonian of of $N$ fully polarized dipoles in two dimensions, written in
dipolar units, is given by
\begin{equation}
  H = -\frac{1}{2}\sum_{i=1}^{N}\nabla_i^2+\sum_{i<j}\frac{1-3\lambda^2\cos^2
  \theta_{ij}}{r_{ij}^3} \ ,
  \label{Manybody_Sch_equation}
\end{equation}
where $r_{ij}$ and $\theta_{ij}$ are the relative distance and polar angle 
formed by dipoles $i$ and $j$, respectively. 
This Hamiltonian is valid only when Van der Waals interaction 
can be neglected in front of dipolar forces. 
As we have commented in the previous Section, we restrict our study to 
polarization angles $\alpha<\alpha_c=0.615$, thus ensuring the potential 
is fully repulsive and the system is stable.

In order to guide properly the diffusion process and improve the variance
of the results one introduces in the DMC method a trial wave function for 
importance sampling. In the present case,  we 
use a variational Jastrow wave function of the form
\begin{equation}
  \Psi(\textbf{r}_1,...,\textbf{r}_N) = \prod_{i<j}f(\textbf{r}_{ij})  \ ,
  \label{Jastrow_wf}
\end{equation} 
where $\textbf{r}_{ij}=\textbf{r}_i-\textbf{r}_j$. 
Notice that due to the
anisotropy of the system, the two-body correlation factor $f(\textbf{r})$
depends on the whole vector, rather than on its magnitude only.  If the
density is not high,  the low-energy two-body scattering solution greatly influences
the  properties of the many body system, as three body scattering processes
have extremely low probability. For this reason, we use as a Jastrow 
two-body correlation function the anisotropic  zero-energy
scattering solution~\cite{Macia_2011} matched at some 
large distance, $\xi$,  with the
symmetrized form of a phononic wave function $f_{\text{ph}}(r_{ij}) =
exp(-C/r_{ij})$,\cite{Reatto_1967} $\xi$ being a variational parameter.

\begin{figure}
  \centering
  \includegraphics[width=0.8\linewidth]{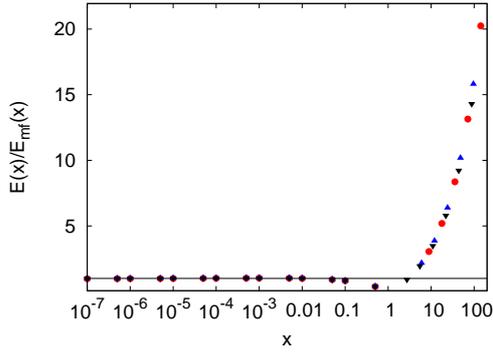}
  \caption{(Color online) Ratio of the energy per 
    particle of the gas of dipoles for
    different polarization angles to the mean field
    prediction (\ref{E_univ}) as a function of the gas parameter $x=n a^2$. 
    Red circles, blue triangles and black inverted
    triangles correspond to $\alpha=0.2, 0.4$ and $0.6$,
    respectively.  The line corresponds to the mean field prediction.}
  \label{E_univ_fig}
\end{figure}

Figure \ref{E_univ_fig} shows the ratio of the energy per particle obtained
from our DMC  calculations and the mean field prediction \cite{Schick_1971}
\begin{equation}
  \frac{E_{MF}}{N} = \frac{\hbar^2}{2ma^2}\frac{4\pi x}{|\ln x|} \ ,
  \label{E_univ}
\end{equation}
where $x= n a^2$ is the gas parameter, with $n$ and $a$ being the density and the
s-wave scattering length, respectively. We report results for three polarization 
angles  $\alpha=0.2, 0.4, 0.6$.
It is important to notice that, for a given value of the gas parameter $x$, 
different polarization angles imply different scattering length
values and different  densities. However, and as it can be seen from the
figure, all energies corresponding to the same $x$ collapse into the same
curve with very small deviations even at the higher densities considered. 
That means that the effects of the anisotropy of the interaction are
accurately contained  in the polarization dependent scattering length
$a(\lambda)$, which is well approximated by the law~\cite{Macia_2011}
\begin{equation}
  a(\lambda) = e^{2\gamma}\left(1-\frac{3\lambda^2}{2}\right) \ ,
  \label{Scattering_length_eq}
\end{equation}
where $\gamma=0.5772...$ is the Euler-Mascheroni constant and the factor
$e^{2\gamma}$ corresponds to the s-wave scattering length of the isotropic
dipolar system in 2D ($\alpha=0$)~\cite{Ticknor_2009}. The scaling law
observed in the energy, as reported in Fig. \ref{E_univ_fig}, may be also
attributed to the small contributions to the energy coming from anisotropic
terms (explicit $\theta$ terms) contained in the local energy estimator.

\begin{figure}
  \centering
  \includegraphics[width=0.9\linewidth]{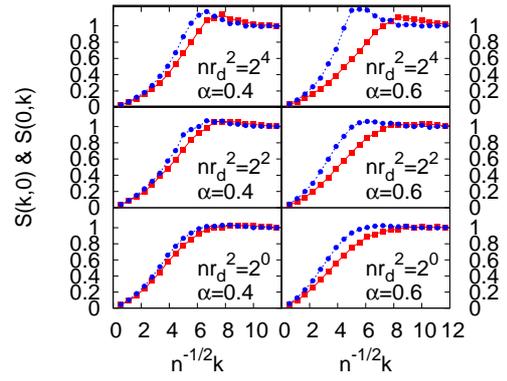}
  \caption{(Color online) Static structure function for polarization
    angles $\alpha=0.4$ and $\alpha=0.6$ for $nr_d^2=2^4, 2$ and 
    $2^{0}$. Red squares and blue circles show the two cuts 
    $S(k,0)$ and $S(0,k)$, respectively.}
  \label{Sk_fig}
\end{figure}

The anisotropic character of the dipole-dipole interaction has a direct
influence on the ground-state wave function that is mirrored in the ground
state expectation values of many-body operators. DMC allows us to evaluate
pure estimations \cite{Casu_1995} of these observables. 
Contrarily to isotropic fluids, the static structure factor $S(\textbf{k})$,
in the dipolar gas depends on the full vector $\textbf{k} = (k_x,k_y)$ rather than on
its magnitude.
Figure \ref{Sk_fig} shows  two cuts of $S(\textbf{k})$ along the perpendicular and 
parallel
directions with respect to the polarization plane, corresponding to the
lines where the interaction is most and least repulsive, respectively. As
expected, the effect of the anisotropy is more clearly seen at high
densities and large polarization angles. In particular, for fixed
$\alpha$
the separation between $S(k,0)$ and $S(0,k)$ is enhanced with increasing
density. In much the same way the separation between the  two cuts of the
structure factor also increases for fixed density when the polarization
angle is  increased. At the largest densities considered, the
effect of increasing the polarization angle makes the peak in
$S(\textbf{k})$ increase in the strongly interacting direction and
decrease  in the weakly interacting one. This behavior is clearly observed in
the Figure for the case $nr_d^2=2^4$ and $\alpha=0.6$.

\begin{figure}
  \centering
  \includegraphics[width=0.9\linewidth]{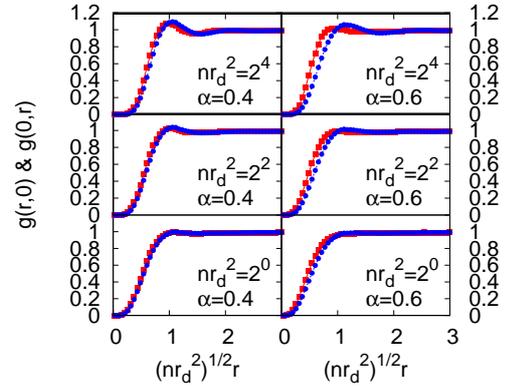}
  \caption{(Color online) Two-body distribution function for polarization
    angles $\alpha=0.4$ and $\alpha=0.6$ for $nr_d^2=2^4, 2$ and 
    $2^{0}$. Red squares and blue circles show the two cuts 
    $g(r,0)$ and $g(0,r)$, respectively.}
  \label{gr_fig}
\end{figure} 

The dependence of the interaction potential on the polar angle $\theta$
induces also changes in the distribution functions in different directions.
We have calculated the two-body distribution function $g(\textbf{r})$ of
the gas for different densities and polarization angles to visualize the
effects of anisotropy in the spatial structure. As in the case of the
static structure function, we have selected two cuts, $g(r,0)$ and
$g(0,r)$,
corresponding to the less and more repulsive directions. Selected results
for the same densities and angles reported in Fig. \ref{Sk_fig} are shown 
in Fig. \ref{gr_fig}. As one can see, the anisotropy is also observed in
the spatial structure but the effect is smaller than the one observed in 
$S(\textbf{k})$. Nevertheless, looking at the case
($nr_d^2=2^4$,$\alpha=0.6$) one observes that $g(r,0)$ 
has less structure than $g(0,r)$ which  has a well defined first peak.

\section{Excitation spectrum}

A relevant issue in the study of tilted dipolar gases is the influence of
the anisotropy of the interaction on the collective excitation spectrum. In
this Section, we analyze this problem within two standard methods
used currently in the study of Bose fluids: the Feynman and Bogoliubov
approximations.

The Feynman spectrum is easy to derive from a simple sum rules argument and
provides a single line in $(\textbf{k},\omega)$ space corresponding to a
set of infinite lifetime excitations of energy \cite{Feynman_1954}
\begin{equation}
  \epsilon(\textbf{k}) = \frac{\hbar^2k^2}{2mS(\textbf{k})} \ .
  \label{Feynman_spectrum}
\end{equation}
In this approximation, $\epsilon(\textbf{k})$ depends directly on the static
structure factor,  the only non-trivial quantity, and provides an upper
bound to the actual excitation spectrum \cite{Boro_1995}. In systems like
liquid $^4$He, this bound is closer to the experimental mode the  lower the
total momentum is.

On the other hand, we can study the excitation spectrum of the low density
two-dimensional dipolar gas in the framework of the mean-field theory using
the 2D time-dependent Gross-Pitaevskii equation, 
\begin{equation}
  i\hbar\frac{\partial \psi}{\partial t} = -\frac{\hbar^2}{2m}\nabla^2\psi
  + g|\psi|^2\psi \ ,
  \label{GP_equation}
\end{equation} 
where $g$ is the 2D coupling constant $g =
\frac{4\pi\hbar^2}{m}\frac{1}{|\log na^2|}$~\cite{Schick_1971}. Performing a standard
Bogoliubov-deGennes linearization one finds the well-known Bogoiubov spectrum 
\begin{equation}
  \epsilon(\textbf{k}) =
  \sqrt{\frac{\hbar^2k^2}{2m}\left(\frac{\hbar^2k^2}{2m}+2gn\right)} \ .
  \label{spectrum_bog}
\end{equation}
Although the spectrum obtained using this approach
has contributions coming from the anisotropic character of the interaction
due to the polarization angle dependence  of the 
scattering length, not all contributions of the same order are taken into
account. This simple  Bogoliubov approach disregards the contribution
coming from higher angular momentum channels, keeping only s-wave
scattering processes. However, we know that different angular momentum
channels couple in a  non-trivial way in a dipolar system and so we have to
take them into account. We know from the
analysis of the zero-energy two-body problem that higher order  partial
wave contributions appear with higher orders in $\lambda^2$, so the leading
corrections  appear in $d$-wave. In order to consider the contribution of
the d-wave we use the following pseudo-potential
\begin{equation}
  V_{ps}(\textbf{r}) = g\delta^{(2)}(\textbf{r})-\frac{C_{dd}}{4\pi}
  \frac{3\lambda^2\cos 2\theta}{r^3}
  \label{GP_pseudopotential}
\end{equation} 
that leads to the following Gross-Pitaevskii equation
\begin{equation}
  \begin{split}
    &i\hbar\frac{\partial \psi}{\partial t} = -\frac{\hbar^2}{2m}\nabla^2\psi + g|\psi|^2\psi-\\
    \frac{C_{dd}}{4\pi}&\left(\int d\textbf{x}'\frac{3\lambda^2\cos 2\theta}{|\textbf{x}
      -\textbf{x}'|^3}|
    \psi(\textbf{x}',t)|^2\right)\psi \ .
  \end{split}
  \label{GP_equation_mod}
\end{equation}
The functional form of the pseudopotential $V_{ps}(\textbf{r})$
(\ref{GP_pseudopotential}) as a sum of two terms, one isotropic and another
anisotropic, follows the same prescription used in the three-dimensional
problem~\cite{Yi_2000,Derevianko_2003}.   
One can consider a linear perturbation of the condensate wave function of
the system of the form
\begin{equation}
  \psi(\textbf{x},t) = e^{-\frac{i}{\hbar}\mu
  t}(\sqrt{n}+\delta\psi(\textbf{x},t)) \ ,
  \label{GP_wavefun}
\end{equation}
where the perturbative term $\delta\psi(\textbf{x},t)$ is given by
\begin{equation}
  \delta\psi(\textbf{x},t) = c e^{i(\textbf{k}\cdot\textbf{x}-\omega t)} \ ,
\end{equation}
where $c$ is the (small) perturbation amplitude.

\begin{figure}
  \centering
  \includegraphics[width=0.9\linewidth]{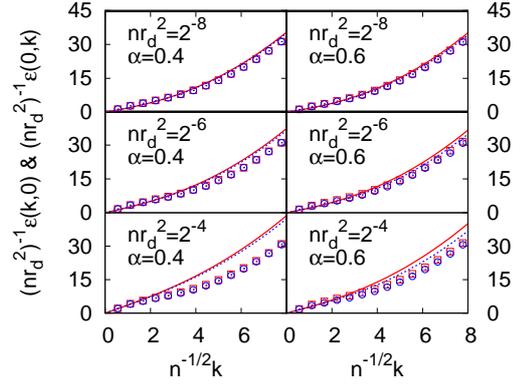}
  \caption{(Color online) Comparison of Feynman (symbols) and Bogoliubov
    (lines) excitation spectrum for angles $\alpha=0.4$ and $\alpha=0.6$
    and $nr_0^2=2^{-4}, 2^{-6}$ and $2^{-8}$. Red solid  and blue dashed
    curves show the two cuts $\epsilon(k,0)$ and $\epsilon(0,k)$
    corresponding to Bogoliubov approximation respectively. Red open
    squares and Blue open circles show  $\epsilon(k,0)$ and $\epsilon(0,k)$
    corresponding to Feynman approximation respectively}
  \label{Ek_bog_fig}
\end{figure}

By inserting (\ref{GP_wavefun}) into Eq. (\ref{GP_equation_mod}), and
neglecting non-linear terms,  one finds the equation fulfilled by the small
perturbation $\delta\psi$,
\begin{equation}
  \begin{split}
  i\hbar\frac{\partial \delta\psi}{\partial t} = & -\frac{\hbar^2}{2m}\nabla^2\delta\psi +
   (2gn-\mu)\delta\psi + gn\delta\psi^*+\\
   &n(F(\textbf{k})\delta\psi^*+F(-\textbf{k})\delta\psi) \ ,
  \end{split}
\end{equation}
where $F(\textbf{k})$ is given by
\begin{equation}
  F(\textbf{k}) = \frac{C_{dd}}{4\pi}
  \left(\int d\textbf{y}\frac{3\lambda^2\cos 2\theta}{|\textbf{y}|^3}
  e^{i\textbf{k}\cdot\textbf{y}}\right) ,
\end{equation}
and $\textbf{y}=\textbf{r}-\textbf{r}'$. Now, taking into account that for
a dilute homogeneous system the chemical potential is $\mu=gn$ and  that for the
two-dimensional dipole-dipole interaction $F(\textbf{k}) = F(-\textbf{k})$,
we  finally arrive at the following expression for the Bogoliubov spectrum
\begin{equation}
  \epsilon(\textbf{k})=
  \sqrt{\frac{\hbar^2k^2}{2m}\left(\frac{\hbar^2k^2}{2m}+2n(g+\pi k 
  \lambda^2 \cos 2\theta_{\textbf{k}})\right)} \ ,
\end{equation}
where $\theta_{\textbf{k}}$ is the angle formed by the momentum of 
the excitation and the $x$-axis.
 
 \begin{figure}
  \centering
  \includegraphics[width=0.9\linewidth]{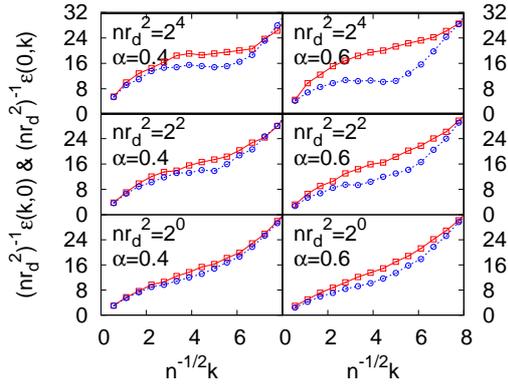}
  \caption{(Color online) Feynman excitation spectrum for
    angles $\alpha=0.4$ and $\alpha=0.6$ for $nr_0^2=2^4, 2$ and 
    $2^{0}$. The red open squares and blue open circles show the two cuts 
    $\epsilon(k,0)$ and $\epsilon(0,k)$, respectively.}
  \label{Ek_fig}
\end{figure} 
  
The comparison between the Bogoliubov approximation gi\-ven in this
expression and the excitation spectrum  obtained from DMC calculations
using the Feynman approximation is shown in Fig. \ref{Ek_bog_fig} for
several  values of the density and polarization angle. We can see from
Fig. \ref{Ek_bog_fig} that,  as expected, the Bogoliubov and Feynman
approximations coincide at very low densities. It is also noticeable the
fact that, for a given value of the density, the Bogoliubov approximation
is  closer to the Feynman prediction at large polarization angles. This
stresses  once again that the relevant quantity describing the low density
dipolar Bose gas is the gas parameter  $x=na^2$, and according to equation
(\ref{Scattering_length_eq}) the scattering length $a(\lambda)$ decreases 
with increasing polarization angle. For a fixed density, $x$
decreases when $\alpha$ increases, and the Feynman prediction gets closer
to the Bogoliubov mode, which is know to successfully characterize the
excitation spectrum of Bose gases when $x \rightarrow 0$.
We can  conclude from Fig.
\ref{Ek_bog_fig} that  Feynman and Bogoliubov approximations are close to
each other at small values of the momentum $k$. Finally, one also sees that
the excitation spectrum becomes isotropic when $k \rightarrow 0$. 

Furthermore, the Bogoliubov approximation is expected to be valid only at very
low densities while the Feynman approximation is known to provide an upper
bound to the exact excitation spectrum of the system. To have some insight
on how $\epsilon(\textbf{k})$ evolves with the density we
show in Figure \ref{Ek_fig} the Feynman mode  at higher
values of $n$.  The results presented in the figure correspond to densities
that are still far from the crystallization  point of the isotropic system
~\cite{Buchler_2007,Astra_2007}. From Fig. \ref{Ek_fig},
one can see that with increasing density the spectrum develops a
roton-like  minimum which for fixed density and polarization angle is
deeper in the most repulsive direction. It is interesting to notice that as
the anisotropy of the interaction is increased,i.e., when the  polarization angle
grows, the roton minimum is deeper in the more repulsive direction while in
the orthogonal direction the spectrum does not show any minimum in the
range of considered densities. In fact, the
emergence of the roton and its eventual zero-energy limit has been
discussed as a clear signature of the instability of the system when the
critical polarization angle is higher than
$\alpha_c$~\cite{Santos_2003,Hufnagl_2011}.

\section{Summary and Conclusions}

To summarize, in this work we have described some properties of a
two-dimensional dipolar Bose gas  where the polarization field forms an
angle $\alpha$ with the normal direction. The projection of  the
polarization vector on the plane defines the x-axis, where the interaction
potential is weaker than in any other direction. In this situation there is
a critical polarization angle $\alpha_c=0.615$  where the potential starts
to show attractive regions.  We
have used the two-body zero energy wave function to build a many-body
Bijl-Jastrow wave function that we used as an input for  our diffusion
Monte Carlo simulations of the homogeneous polarized dipolar Bose gas. We
have presented  results for the energy per particle in comparison with the 
well known mean field prediction of the two
dimensional equation of state at low densities. The scaling of the energy in the gas
parameter is preserved up to values of $x$  much larger than those of the
mean-field regime, corresponding to values $x \leq 10^{-2}$, 
in which the energy only depends on the gas parameter and not on the specific interaction. 

In the second part, we have studied the excitation 
spectrum of the system in two different frameworks, the Bogoliubov and Feynman
approximations. We have  shown that the agreement between  these two
approaches is very good at very low densities as it was expected. We have
derived the first a\-ni\-so\-tro\-pic correction to the Bogoliubov spectrum, and we
have shown that in the range of validity of the Bogoliubov approximation
the anisotropy plays an extremely small role. At higher densities, the
Feynman approximation predicts a strongly a\-ni\-so\-tro\-pic spectrum showing a
roton-like minimum that depends on the direction of the momentum.
 
\begin{acknowledgement}

This work has been partially supported by Grants No.~FIS2011-25275
from DGI (Spain), Grant No.~2009-SGR1003 from the Generalitat de
Catalunya (Spain).

\end{acknowledgement}



\begin{thebibliography}{99}

\bibitem{Griesmaier_2005}
  A. Griesmaier, J. Werner, S.Hensler, J.Stuhler, and T. Pfau, 
  Phys. Rev. Lett.~{\bf 94}, 160401 (2005).

\bibitem{Stuhler_2005} J. Stuhler, A. Griesmaier, T. Koch, M. Fattori,
  T. Pfau, S. Giovanazzi, P. Pedri, and L. Santos,
  Phys. Rev. Lett.~{\bf 95}, 150406 (2005).

\bibitem{KKNi_2008}
  K.-K. Ni, S. Ospelkaus, M. H. G. de Miranda, A. Pe'er, B. Neyenhuis, J. J. Zirbel,
  S. Kotochigova, P. S. Julienne, D. S. Jin, and J. Ye,
  Science~{\bf 322}, 231 (2008).

\bibitem{Ospelkaus_2009}
  S. Ospelkaus, A. Pe'er, K.~K. Ni, J. J. Zirbel, B. Neyenhuis, S. Kotochigova, 
  P. S. Julienne, J. Ye, and D. S. Jin, 
  Nature Phys.~{\bf 4}, 622 (2009).

\bibitem{KKNi_2010} K.-K. Ni, S. Ospelkaus, D. Wang, G. Qu\'em\'ener,
  B. Neyenhuis, M. H. G. de Miranda, J. L. Bohn, J. Ye, and  D. S. Jin,
  Nature~{\bf 464}, 1324 (2010).

\bibitem{Lu_2011} M. Lu, N. Q. Burdick, S. H. Youn and B. L. Lev, 
  Phys. Rev. Lett. \textbf{107}, 19041 (2011).

\bibitem{Lu_2012} M. Lu, N. Q. Burdick and B. L. Lev, 
  Phys. Rev. Lett. \textbf{108} 215301 (2012).

\bibitem{Aikawa_2012} K. Aikawa, A. Frisch, M. Mark, S. Baier, A. Rietzler,
  R. Grimm and F. Ferlaino, 
  Phys. Rev. Lett. \textbf{108} 210401 (2012).

\bibitem{Santos_2000}
  L. Santos, G. V. Shlyapnikov, P. Zoller and M. Lewenstein, 
  Phys. Rev. Lett. \textbf{85} 1791 (2000).

\bibitem{Santos_2003}
  L. Santos, G. V. Shlyapnikov and M. Lewenstein, 
  Phys. Rev. Lett. \textbf{90} 250403 (2003).

\bibitem{Ticknor_2009}
  C. Ticknor, Phys. Rev. A \textbf{80}, 052702 (2009).

\bibitem{Ticknor_2010}
  C. Ticknor, Phys. Rev. A \textbf{81}, 042708 (2010).

\bibitem{Ticknor_2011}
  C. Ticknor, Phys. Rev. A \textbf{84}, 032702 (2011).

\bibitem{Cai_2010}
  Y. Cai, M. Rosenkranz, Z. Lei and W. Bao, 
  Phys. Rev. A \textbf{82}, 043623 (2010).

\bibitem{Buchler_2007}
  H. P. Buchler, E. Demler, M. Lukin, A. Micheli, N. Prokof'ev, G. Pupillo and P. Zoller,
  Phys. Rev. Lett. \textbf{98}, 060404 (2007).

\bibitem{Astra_2007}
  G. E. Astrakharchik, J. Boronat, I. L. Kurbakov and Y. E. Lozovik,
  Phys. Rev. Lett. \textbf{98}, 060405 (2007).

\bibitem{Astra_2009}
  G. E. Astrakharchik, J. Boronat, J. Casulleras, I. L. Kurbakov and Y. E. Lozovik,
  Phys. Rev. A \textbf{79}, 051602 (2009).

\bibitem{Filinov_2010}
  A. Filinov, N. V. Prokof'ev and M. Bonitz,
  Phys. Rev. Lett. \textbf{105}, 070401 (2010).

\bibitem{Macia_2011}
  A. Macia, F. Mazzanti, J. Boronat, and R. E. Zillich, 
  Phys. Rev. A~{\bf 84}, 033625 (2011).

\bibitem{Mazzanti_2009}
  F. Mazzanti, R. E. Zillich, G. E. Astrakharchik, and J. Boronat,
  Phys. Rev. Lett.~{\bf 102}, 110405 (2009).
  
\bibitem{Ticknor_2010b}
  C. Ticknor, R. M. Wilson, and J. L. Bohn, 
  Phys. Rev. Lett.~{\bf  106}, 065301 (2011).

\bibitem{Hufnagl_2011}
  D. Hufnagl, R. Kaltseis, V. Apaja and R. E. Zillich,
  Phys. Rev. Lett. \textbf{107}, 065303 (2011).

\bibitem{Feynman_1954} 
  R. P. Feynman, Phys. Rev. \textbf{94}, 262 (1954).

\bibitem{Reatto_1967}
  L. Reatto and G. V. Chester, Phys. Rev.~{\bf 155}, 88 (1967).

\bibitem{Schick_1971}
      M. Schick, Phys. Rev. A \textbf{3}, 1067 (1971).

\bibitem{Casu_1995}
  J. Casulleras and J. Boronat, Phys. Rev. B~{\bf 52}, 3654 (1995).

\bibitem{Boro_1995}
  J. Boronat, J. Casulleras, F. Dalfovo, S. Stringari and S. Moroni, 
  Phys. Rev. B \textbf{52}, 1236 (1995).

\bibitem{Yi_2000}
    S. Yi and L. You, Phys. rev. A {\bf 63}, 053607 (2000).
    
\bibitem{Derevianko_2003}    
    A. Derevianko, Phys. Rev. A {\bf 67}, 033607 (2003). 
  
\end{thebibliography}
\end{document}